\documentstyle[wrapfig,epsfig]{aipproc}

\newcommand\etal{{\it et al.}}
\begin{document}
\title{Simulations of Astrophysical Fluid Instabilities
}
\begin{small}
\vspace{-0.4in}
\noindent
(This manuscript will appear in the proceedings of the 
20th Texas Symposium on Relativistic Astrophysics.)
\end{small}

\author{A. C. Calder$^{*,\dagger}$, B. Fryxell$^{*,\ddagger}$, 
R. Rosner$^{*,\dagger,\ddagger}$, L. J. Dursi$^{*,\dagger}$,  \\
K. Olson$^{*,\ddagger,\S}$, P. M. Ricker$^{*,\dagger}$, 
F. X. Timmes$^{*,\dagger}$, M. Zingale$^{*,\dagger}$,  \\
P. MacNeice$^{\S}$, and H. M. Tufo$^{*}$
}

\address{$^*$Center for Astrophysical Thermonuclear Flashes\thanks{This work is 
supported by the U.S. Department of Energy under Grant 
No. B341495 to the Center for 
Astrophysical Thermonuclear Flashes at the University of Chicago.},
University of Chicago  Chicago, IL 60637 \\
$^{\dagger}$Department of Astronomy and Astrophysics,
University of Chicago  
Chicago, IL 60637 \\
$^{\ddagger}$ Enrico Fermi Institute,
                 The University of Chicago,
                 Chicago, IL  60637 \\
$^{\S}$NASA Goddard Space Flight Center,
                 Greenbelt, MD  20771
}

\maketitle

\begin{abstract}
We present direct numerical simulations of mixing at Rayleigh-Taylor 
unstable interfaces performed with the FLASH code, developed at 
the ASCI/Alliances Center for Astrophysical Thermonuclear Flashes at 
the University of Chicago. We present initial results of single-mode studies
in two and three dimensions. Our results indicate that three-dimensional instabilities
grow significantly faster than two-dimensional instabilities and that grid
resolution can have a significant effect on instability growth rates.
We also find that unphysical diffusive mixing occurs at the
fluid interface, particularly in poorly resolved simulations.
\end{abstract}

\section*{Introduction}

\vspace{-0.06in}  
Many of the problems of interest in relativistic astrophysics
involve fluid instabilities. The shock of a core-collapse supernova 
propagating through the outer layers of the collapsing star, for example, 
is subject to Rayleigh-Taylor instabilities
occurring at the boundaries of the layers. 
A fluid interface is said to be Rayleigh-Taylor 
unstable if either the system 
is accelerated in a direction perpendicular to the interface such that the 
acceleration opposes the density gradient or if the pressure gradient 
opposes the density gradient \cite{taylor50,chandra61}.
Growth of these instabilities
can lead to mixing of the layers. The early observation
of $^{56}$Co, an element formed in the core, in SN 1987A 
strongly suggested that mixing 
did indeed play a fundamental role in the dynamics. Following this observation,
supernova modelers embraced multi-dimensional models with the goal of 
understanding the role of fluid instabilities in the core collapse supernova 
process \cite{arnett89}.
Despite years of modeling these events, many
fundamental questions remain concerning 
fluid instabilities and mixing.
In this manuscript, we present early results of our research into resolving 
fluid instabilities with FLASH, our simulation code for astrophysical reactive
flows. 

The FLASH code \cite{fryxell00} 
is an adaptive mesh, parallel simulation code for studying
multi-dimensional compressible reactive flows in astrophysical environments. 
It uses a customized version of the
PARAMESH library \cite{macneice00} to manage a block-structured
adaptive grid, placing resolution elements only where needed in order
to track flow features. FLASH solves the compressible Euler equations 
by an explicit, directionally split version of the
piecewise-parabolic method \cite{colella84} and allows
for general equations of state using the method of Colella \& Glaz \cite{colella85}. 
FLASH solves a separate advection equation for the partial density of each 
chemical or nuclear species as required for reactive flows. The code does
not explicitly track interfaces between fluids, so a small amount
of numerical mixing can be expected during the course of a calculation.
FLASH is implemented in Fortran 90 and
uses the Message-Passing Interface library to achieve portability.
Further details concerning the algorithms used in the code, the structure of 
the code, verification tests, and performance may be found
in Fryxell \etal \cite{fryxell00} and Calder \etal \cite{calder00}.

\section*{Results}

From our single-mode Rayleigh-Taylor studies, we find significantly faster
instability growth rates in three-dimensional simulations
than in two-dimensional simulations.
In addition, we find that 
obtaining a converged growth rate requires at least 25 grid points per 
wavelength of the perturbation, that grid noise seeds small scale 
structure, and that the 
amount of small scale structure increases with resolution 
due to the lack of a physical dissipation mechanism (such as a viscosity).
Another result is that poorly-resolved simulations exhibit a 
significant unphysical diffusive mixing. Figure 1 shows the
growth of bubble and spike amplitudes for two well-resolved 
simulations beginning from equivalent initial conditions. 
The three-dimensional result (left panel) shows faster growth than the 
two-dimensional result (right panel).
Results of our single-mode studies will appear in Calder
\etal \cite{calder01}. 
\begin{figure}[h] 
\centerline{\epsfig{file=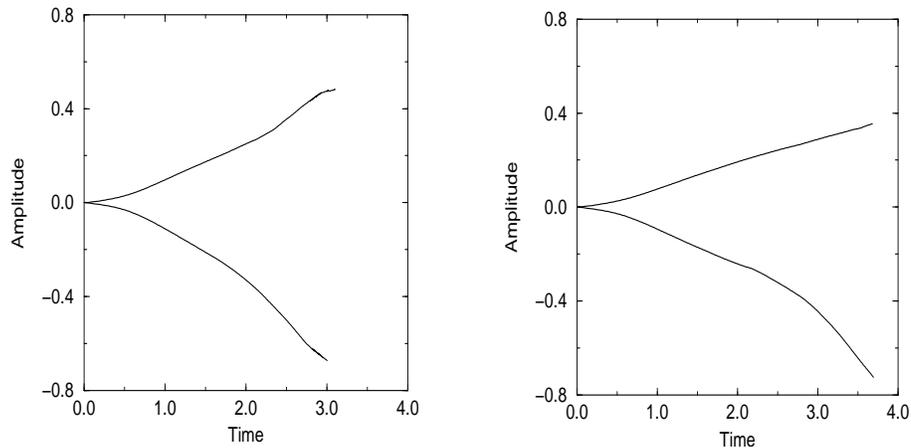,height=2.3in,width=4.7in}}
\vspace{10pt}
\caption{
Bubble and spike amplitudes for two-dimensional (right) and three-dimensional (left) simulations
of single-mode instabilities. The resolutions are 128 X 768 (2-d) and 128 X 128 X 768 (3-d). 
The amplitudes are measured by tracking the advection of each fluid.
The initial conditions consisted of 
a dense fluid ($\rho = 2$) over a lighter fluid ($\rho = 1$) and
$g = 1$. The
initial perturbation consisted of a sinusoidal vertical velocity perturbation of 
2.5\% of the local sound speed with the horizontal components chosen
so the initial velocity field was divergence-free.
}
\end{figure}

Our single-mode studies serve as a prelude
to multi-mode studies, which are works in progress; our single-mode results strongly
suggest that using sufficient resolution is essential in order to obtain
physically-sensible results for these calculations. In the multi-mode case,
bubble and spike mergers
are thought to lead to an instability growth
according to a $t^2$ scaling law, which for the case of a dense 
fluid over a lighter fluid in a gravitational field may be
written as \cite{youngs94}
\begin{equation}  \label{eq:tsq}
h_{b,s} = \alpha_{b,s}gAt^2 
\end{equation}
where $h_{b,s}$ is the height of a bubble or spike, $g$ is the
acceleration due to gravity, $A = (\rho_2 - \rho_1)/(\rho_2 + \rho_1)$ is
the Atwood number
where $\rho_{1,2}$ is the density of the lighter (heavier) fluid,
and $t$ is the time. $\alpha$ is a proportionality `constant' that
may be thought of as a measure of the efficiency of potential energy
release. Experiments and simulations indicate that $\alpha$ lies in
the range of 0.03 to 0.06, and it is thought to depend on Atwood 
number, evolution time, initial conditions, and dimensionality.
See Young \etal\/ \cite{young00} and references therein for a
discussion of experimental results.
Results of our multi-mode studies will appear in
publications of the Alpha Group, a consortium formed by Guy Dimonte in 1998
to determine if the $t^2$ scaling law holds for the growth of
the Rayleigh-Taylor instability mixing layer, and if so, to
determine the value of $\alpha$ 
\cite{dimonte01}.


\begin{references}

\bibitem{taylor50} Taylor, G., {\it Proc. Roy. Soc.}, {\bf A 201}, 192 (1950)

\bibitem{chandra61}Chandrasekhar, S., {\it Hydrodynamic and Hydromagnetic Stability}, 
New York: Dover, 1961, ch. X, pp. 428-480.

\bibitem{arnett89} Arnett, D., Fryxell, B., and M\"{u}ller, E., {\it Ap. J.}, {\bf 341}, L63 (1989)

\bibitem{fryxell00}Fryxell, B. A., \etal, {\it Ap. J. S.} {\bf 131}, 273 (2000)

\bibitem{macneice00} MacNeice, P., \etal, {\it Comp. Phys. Comm.}, {\bf 126}, 330 (2000)

\bibitem{colella84}Colella, P. and  Woodward, P., {\it J. Comp. Phys.} {\bf 54}, 174 (1984)

\bibitem{colella85} Colella, P. and  Glaz, H. M.,  {\it J. Comp. Phys.} {\bf 59}, 264 (1985)

\bibitem{calder00} Calder, A. C., \etal, in Proc. Supercomputing 2000, 
IEEE Computer Soc., 2000 

\bibitem{calder01} Calder, A. C. \etal, in prep. (2001)

\bibitem{youngs94}Youngs, D. L., {\it Lasers and Particle Beams}, {\bf  12}, no. 4, 725 (1994)

\bibitem{young00}Young, Y.-N., \etal, {\it J. Fluid Mech.}, in press (2001)

\bibitem{dimonte01} Dimonte, G. \etal, in prep. (2001)

\end{references}
\end{document}